\documentclass[prd, twocolumn, nofootinbib, amsmath, amssymb,showpacs]{revtex4}

\usepackage{graphicx,epsfig,psfig}
\usepackage{dcolumn}
\usepackage{bm}

\begin{document}

\title{On causality and superluminal behavior in classical field theories. Applications to k-essence theories and MOND-like theories of gravity.}

\author{Jean-Philippe Bruneton}
\email{bruneton@iap.fr}
\affiliation{$\mathcal{G}\mathbb{R}\varepsilon\mathbb{C}\mathcal{O}$, Institut d'Astrophysique de Paris, UMR 7095-CNRS, Universit\'e Pierre et Marie Curie - Paris 6, 98 bis boulevard Arago F-75014, Paris, France}

\date{\today}

\begin{abstract}
Field theories whose full action is Lorentz invariant (or
diffeomorphism invariant) can exhibit superluminal behaviors
through the breaking of local Lorentz invariance. Quantum induced
superluminal velocities are well-known examples of this effect.
The issue of the causal behavior of such propagations is somewhat
controversial in the literature and we intend to clarify it. We
provide a careful analysis of the meaning of causality in
classical relativistic field theories, and we stress the role
played by the Cauchy problem and the notions of chronology and
time arrow. We show that superluminal behavior threaten causality
only if a prior chronology on spacetime is chosen. In the case
where superluminal propagations occur, however, there is at least
two non conformally related metrics on spacetime and thus two
available notions of chronology. These two chronologies are on
equal footing and it would thus be misleading to choose \textit{ab
initio} one of them to define causality. Rather, we provide a
formulation of causality in which no prior chronology is assumed.
We argue this is the only way to deal with the issue of causality
in the case where some degrees of freedom propagate faster than
others. We actually show that superluminal propagations do not
threaten causality. As an illustration of these conceptual issues,
we consider two field theories, namely k-essences scalar fields
and bimetric theories of gravity, and we derive the conditions
imposed by causality. We discuss various applications such as the
dark energy problem, MOND-like theories of gravity and varying
speed of light theories.
\end{abstract}


\maketitle

\section{Introduction}

The question of the causal behavior of superluminal propagations
has been often debated, in rather different contexts: tachyonic
particles (e.g. \cite{Tolman,
antitelephone,LiberatiVisser,DolgovAdetruire,DolgovDetruit}),
field theories
\cite{Susskind,Arkani-Hamed,Lim,Bonvin,Cerenkov,BekensteinDisformal,BekensteinDisformalApplication,Bekenstein2004}
and quantum induced superluminal propagations
\cite{LiberatiVisser,FullList,Shore, Scharnhorst}. The issue is
however controversial, even on specific contexts, and we intend to
clarify it.

We will focus on classical field theories \textit{whose full
action is Lorentz invariant} (or diffeomorphism invariant whenever
Einstein's gravity is taken into account). In that case,
superluminal behavior can only arises from spontaneous breaking of
local Lorentz invariance by non trivial backgrounds (or vacua).
Superluminal behavior then occurs when some sector of the theory
is left unbroken, so that some degrees of freedom can propagate
faster than other ones \cite{Moffat92}. Superluminal propagations
induced by vacuum polarization provide well-known examples of this
phenomenon.

Note that we define superluminal behavior as going faster than
gravitons [i.e., gravitational waves]. We denote $c$ the speed of
gravitons in vacuum. This maybe unconventional definition will not
affect our arguments. In the standard theory of gravity photons
propagate along the gravitational metric so that this definition
is the usual one in that case.

Our general analysis will enable us to investigate the causal
behavior of any field coupled to standard gravity. We will
illustrate our arguments in three important cases: k-essence
scalar fields, bimetric theories of gravity and quantum induced
superluminal propagations. We will make constant use of many
important ideas and tools that were developed in order to
investigate the causal behavior of GR, and we refer notably to
\cite{Hawking, Wald}.

In Sec.~\ref{Causalite}, we analyze in detail the meaning of
causality in classical field theories. We stress the important
role played by the notions of chronology, time arrow and Cauchy
problem in its definition. As an illustration, we show how
causality is usually formulated in the theory of general
relativity (GR). We stress that superluminal propagations are
automatically discarded if one assumes that causality should be
defined with respect to the chronology induced by the
gravitational metric. This \textit{postulate} actually ``sets the
[gravitational] metric apart from the other fields on
$\mathcal{M}$ and gives it its distinctive geometrical character''
\cite{Hawking}. This prior assumption cannot be supported by any
mathematical reason but only by experiment. If superluminal
propagations were found in the laboratory, it would thus
\textit{not} mean that causality, locality, or the entire
framework of special relativity would be lost, but simply that the
gravitational field does not have any fundamental geochronological
character. In particular, causality should not be expressed in
terms of the chronology induced by the gravitational metric field.
As a consequence, it is necessary to drop such prior assumptions
to address the issue of the causal behavior of superluminal
fields. In Sec.~\ref{CausalityAndSuperluminalBehav} we thus look
for a minimal expression of causality in which no prior chronology
is assumed. In that framework, superluminal propagations are
generically allowed.

Before moving to superluminal and causal theories, we display two
well known examples of noncausal theories and notably the
tachyonic one. It enables us to stress how noncausal behaviors are
related to constraints on initial data. We also comment on the
causal paradoxes that arise in a non causal theory. We finally
briefly investigate what kind of field theory can lead to
superluminal propagation.

We consider in Sec.~\ref{SectionKessence} a k-essence scalar field
$\varphi$, which can propagate superluminally along an effective
metric $G^{\mu\nu}[\varphi]$. We find that the scalar field is
causal provided that the spacetime embedded with the effective
metric $G^{\mu\nu}_0$ is globally hyperbolic. Here $G^{\mu\nu}_0$
is the effective metric evaluated on a solution $\varphi_0$ to the
field equation (the background). This condition puts some generic
constraints on the free function defining the theory, and is
generally satisfied on reasonable backgrounds. We emphasize that
the claim that such scalar fields are not causal even on globally
hyperbolic backgrounds
\cite{Susskind,Arkani-Hamed,Bekenstein2004,BekensteinMilgrom84}
can only hold if a prior role is attributed to the gravitational
(or flat) metric, an assumption that, we believe, was implicit in
these works. The only threat for causality actually lies in global
properties of backgrounds \cite{Arkani-Hamed}, and this can be
related to the so-called chronology protection conjecture.

In Sec.~\ref{Bimetrictheories}, we investigate the question of
causality in a bimetric theory of gravity, in which the matter
sector is universally coupled to a metric $\tilde{\mathbf{g}}$
that can differ from the Einstein-Hilbert metric $\mathbf{g}$.
These two metrics generally define two distinct ``causal'' cones.
Photons can travel faster than gravitons, and conversely. If
causality is defined with respect to the chronology induced by the
gravitational metric, then causality forces the matter metric
$\tilde{\mathbf{g}}$ to define a cone which coincide or lie within
the gravitational cone, and conversely. These two choices of
chronology have been considered in the literature, thus leading to
opposite requirements on the theory. This will illustrate the fact
that there is no clear reason why a metric, or a chronology,
should be preferred to the other. We argue that the theory is
actually causal if no prior chronology is assumed, but the
solution to the Cauchy problem depends, of course, on the precise
dynamics of the matter sector.

In Sec.~\ref{SectionQED}, we briefly show how our analysis apply
to the case of quantum induced superluminal propagations. This
enables us to conclude that superluminal velocities do not
threaten causality, a point that were still controversial.

Let us emphasize that the question of causality in k-essences or
bimetric theories is not just of academical interest, since these
theories have drawn much attention recently. Quite generally, it
is claimed that causality forces one of the cone to be wider than
the others ones (be it the gravitational or the matter one)
\cite{Susskind, Arkani-Hamed, Lim, Bonvin,
BekensteinDisformal,Bekenstein2004} and this puts some constraints
on the theory. Our motivated definition of causality will however
not support these claims.

Such a k-essence scalar field was notably used as a (dynamical)
dark energy fluid responsible of the late time acceleration of the
Universe \cite{Chiba}, as well as a fluid that drives inflation
\cite{Damour}. It is also used as a new gravitational field in
addition to the metric one in some relativistic theories of
MOdified Newtonian Dynamics (MOND), that are intended to account
for the mass discrepancy at astrophysical scales without invoking
Dark Matter \cite{Milgrom83, BekensteinMilgrom84,Bekenstein2004}.
Some of these models were notably discarded because of the
presence of superluminal propagations. Bimetric theories are also
an essential piece of some recent MOND-like theories
\cite{Sanders97, Bekenstein2004} (see also
\cite{BimetricDrummond}). Moreover, it represents the best
motivated framework that reproduces a Varying Speed of Light
scenario \cite{VSLMoffat,VSLVisser} that may address the problems
of standard cosmology in a rather different way than inflation
does. Indeed some VSL theories, in which the speed of light $c$ is
replaced by a changing velocity $c(t)$ \textit{inside} the
equation of motions \cite{VSLBarrowetal}, however interesting
phenomenologically, are not satisfying theoretically
\cite{UzanEllis,VSLVisser} (the resulting theory cannot be derived
from an action).

In the last section \ref{Applications}, we consider some
applications of our work to these field theories. We investigate
the links between causality and stability in k-essences theories.
We show that (k-essence) ghosts stabilization might suffer from a
serious problem. We also show that, if one reproduces the MOND
phenomenology with the help of a k-essence scalar field, then a
slight modification of Milgrom's law is necessary for the theory
to be causally well-behaved. It leads to a non trivial
modification of the phenomenology in the very low acceleration
regime. We also comment on a simpler theory that reproduces the
MOND phenomenology, using only one scalar field. The initial value
formulation of this theory has still to be checked and we leave it
for further work. We briefly show how this framework can account
for the Pioneer anomaly, and we finally make some comments about
VSL theories.

We use, throughout the paper, the sign conventions of \cite{MTW}
and notably the mostly $+$ signature. The flat metric is denoted
$\boldsymbol{\eta}$ or $\mathbf{\eta}_{\mu\nu}$ in a coordinate
system. We denote $\mathbf{g}$ the gravitational metric field
(which obeys the Einstein equation) and $\tilde{\mathbf{g}}$ the
matter metric to which matter field couples. Unless specified,
indices are moved with the help of the gravitational metric
$\mathbf{g}$.

\section{\label{Causalite}Causality and superluminal behavior}

In the present paper, we are interested in the superluminal
behavior of some matter fields in a GR-like context. The metric
field $\mathbf{g}$ follows the dynamics induced by the
Einstein-Hilbert action. By spacetime we will always mean a couple
($\mathcal{M},\mathbf{h}$), where $\mathcal{M}$ is a
four-dimensional differentiable manifold and $\mathbf{h}$ some non
degenerate Lorentzian metric on it. A superluminal signal,
\textit{by definition}, propagates along spacelike curves of the
gravitational metric $\mathbf{g}$.

\subsection{\label{CausaChrono}Causality, chronology and the flow of time}

By causality we usually mean the ability to find a cause to an
effect. Since cause and effects are both described in terms of
some physical variables, the usual principle of causality states
that physical variables should be unambiguously determined at a
given time from their values at a time before. We also demand that
this retrodiction can be converted forward in time, so that we
should also be able to predict a future situation from a present
one. The principle of causality thus requires that determinism
holds ``in both directions of time''. Causality demands the
existence and uniqueness of solutions to the equations of motion
given some initial data. In mathematical words, equations of
motion must have a well-posed Cauchy problem, or initial value
formulation.

Note that this definition shows that a time-ordering, or
chronology, must exist between two spacetime points that are
causally connected. Unlike Newton's theory where a global
chronology preexists to dynamics because of the prior topology
assigned to spacetime, namely $\mathbb{R}^3 \times \mathbb{R}$,
field theories (including Einstein's gravity) do \textit{not}
involve a preexistent notion of time and chronology. On the
contrary, any relativistic field $\psi_i$ defines its own
chronology on $\mathcal{M}$ by means of the metric $\mathbf{h}_i$
along which it propagates. Any Lorentzian metric indeed induces a
local chronology in the tangent space through the usual special
relativistic notions of absolute (i.e., Lorentz invariant) future
and past.

Let us consider one particular field $\psi$ that propagates along
a metric $\mathbf{h}$. It is important to note that this metric
may not induce a global chronology on the whole spacetime. This is
the case if local causal cones of $\mathbf{h}$ are distributed on
the manifold in such a way that there exists a curve over
$\mathcal{M}$ which is everywhere future-directed timelike but
closed. In that case, the field $\psi$ propagates along a closed
curve and an event could be \textit{both} the cause and the effect
of another event. Causality requires that it does not happen. The
strongest way to prevent it is to require the global hyperbolicity
\footnote{A globally hyperbolic spacetime is such that a Cauchy
surface $\Sigma$ exists \cite{Wald}. It can then be proved that
the spacetime has the topology of $\Sigma \times \mathbb{R}$, so
that a global time function, i.e., a global chronology exists.} of
the spacetime ($\mathcal{M},\mathbf{h}$).

Let us emphasize that causality also requires a notion of time
flow. It is indeed worth recalling that even if the above metric
$\mathbf{h}$ defines a global notion of future and past, we have
also to require that the field $\psi$ can only propagate in the
future. If it could, on the contrary, propagate both in future and
in past, one could always form a closed timelike curve with it. On
the other hand, global hyperbolicity is enough to guarantee that
no closed \textit{future-directed} timelike curves (CFTC) exists.

When there is a finite number of metrics $\mathbf{h}_i$ on
$\mathcal{M}$, the discussion of global properties is slightly
more involved, see Sec.~\ref{CausalityAndSuperluminalBehav}.

\subsection{\label{CausaliteGR}Causality in general relativity}

Let us illustrate the previous discussion. In GR, causality is
generally expressed by the following properties
\cite{Hawking,Wald}:
\begin{enumerate}
\item[(a)] The gravitational metric $\mathbf{g}$ must define a
global chronology on spacetime. \item[(b)] ``The null cones of the
matter equations coincide or lie within the null cone of the
spacetime metric $\mathbf{g}$''. \cite{Hawking} \item[(c)] The
whole set of equations of motion must admit a well-posed Cauchy
problem.
\end{enumerate}

The point (a) guarantees the existence of a global time ordering
on $\mathcal{M}$ and the point (c) is the formal expression of
determinism. Note that it is a non trivial mathematical property
of systems of differential equations. Its satisfaction in GR
therefore critically depends on the precise dynamics of the matter
sector.

The point (b) excludes faster-than-graviton propagation.
Equivalently then, it means that any initial data set on spacelike
hypersurfaces with respect to the gravitational metric
$\mathbf{g}$ are allowed\footnote{Up to Hamiltonian constraints
arising from various gauge invariance.}. Since the metric
$\mathbf{g}$ defines a global chronology on spacetime, so does any
other metric $\mathbf{h}_i$ associated to the propagation of some
matter fields\footnote{One may indeed easily check that the
following property holds: if ($\mathcal{M},\mathbf{h}$) is
globally hyperbolic and if $\mathbf{h'}$ is a metric whose null
cone coincide or lie within the null cone of the metric
$\mathbf{h}$ \textit{everywhere on} $\mathcal{M}$, then
($\mathcal{M},\mathbf{h'}$) is globally hyperbolic.}. The
conditions (a) and (b) therefore ensures that no CFTC exist for
any metric (be it $\mathbf{g}$ or any metric $\mathbf{h}_i$).

Let us emphasize that, in the above formulation (a)--(c), the
gravitational metric field clearly plays a preferred role. Both
conditions (a) and (b) are such that causality is actually defined
with respect to the chronology induced by the metric $\mathbf{g}$.
This can be understood as a ``\textit{postulate} which sets the
metric \textbf{g} apart from the other fields on $\mathcal{M}$ and
gives it its distinctive geometrical character'' (\cite{Hawking}
$\S 3.2$, emphasis added). The chronology defined by $\mathbf{g}$
is thus assumed to be a preferred one on $\mathcal{M}$.

This postulate is however arguable because, as we pointed out, any
relativistic field induces its own chronology on spacetime. It
notably means that if a signal made of waves of some field
$\psi_i$ propagates between two spacetime points, then these
points can be time-ordered with the help of the metric
$\mathbf{h}_i$, and causality may be preserved even if the field
propagates superluminally. The gravitational metric field
$\mathbf{g}$ is just one particular field on $\mathcal{M}$ and
there is no clear reasons why it should be favored.

Let us finally stress that the three conditions (a)--(c) are
actually not sufficient to guarantee (usual) causality. Indeed, if
information were allowed to travel as well forward as backward in
time, as it is the case in any time reversal invariant theories,
then it would be always possible to have information running along
a closed curve in spacetime. Thus the absence of CFTC does not
suffice to guarantee the usual notion of causality, unless one
specifies that a \textit{flow} of time exists and that information
can only go forward in time. The three points (a)--(c) are clearly
time-reversal invariant statements\footnote{We thank E. Flanagan
(private communication) for a discussion on that point.} and
therefore cannot address this issue. Note that both classical and
quantum field theories are time-reversal invariant so that they
may not be considered as ``causal'' in the usual sense. The
emergence of the flow of time from thermodynamics in quantum field
theory is a quite involved question that we will not consider
here. In the present paper, we will therefore restrict ourselves
to such a time-reversal invariant notion of causality, but one
should keep in mind that thermodynamics may be the key to
understand fully the (coarse-grained) notion of causality.

\subsection{\label{CausalityAndSuperluminalBehav} Causality and superluminal behavior}

We stressed that superluminal behaviors are automatically
discarded by the above postulate. One may try to justify it by
arguing that, since $\mathbf{g}$ reduces locally to
$\boldsymbol{\eta}$ and therefore gives to the tangent space its
special relativistic (Minkowskian) structure, going faster than
gravitons would ``undermine the entire framework of relativity
theory'' \cite{Wald}. Let us however stress that any other metric
$\mathbf{h}_i$ can also be reduced locally to its fundamental form
in adapted ``inertial'' coordinates. These coordinates transform
under the action of the Lorentz group $SO(3,1)$ with an invariant
speed $c_i$ \textit{that can differ from} $c$, if the cones of
$\mathbf{h}_i$ and $\mathbf{g}$ do not coincide [due for instance
to some spontaneous breaking of local Lorentz invariance, see
Sec.~\ref{CausaTheoriesSuperluminiques}]. These coordinates are
actually relevant if one uses rods and clocks made of the field
$\psi$ that propagates along $\mathbf{h}$. The fact that
$\mathbf{g}$ reduces locally to the flat spacetime metric can
therefore not be used to claim that $\mathbf{g}$ should be a
preferred field on $\mathcal{M}$. We will come back to this
important point later, see Sec.~\ref{BimetricSuperlum}.

We may therefore drop the above postulate and look for a minimal
expression of causality in which no prior chronology is assumed.
Let us consider a collection of fields $\psi_i$ that propagate
along some metrics $\mathbf{h}_i$ (gravity included). We are led
to the picture of a finite set of ``causal'' cones at each points
of spacetime. We do not want to prefer one metric with respect to
the others, because there is no reason why, locally, some sets of
coordinates, or some rods and clocks should be preferred:
coordinates are meaningless in GR.

The cones defined by the metrics $\mathbf{h}_i$ may therefore be
in any relative position with respect to one another. Moreover,
these cones may even tip over each others depending on the
location on spacetime. We have however to guarantee that no signal
can propagate along a closed curve. For this it is not sufficient
to require the global hyperbolicity of each spacetime
($\mathcal{M},\mathbf{h}_i$). Indeed, fields may interact
together, and we may form a physical signal which propagates on
spacetime along different metrics.

\begin{figure}
\caption{\label{FigureFuture} The hatched part shows the extended
future defined by two metrics (solid and dashed lines) in the case where one metric defines
a wider cone than the other one (left), and in the opposite case
(right).}
\includegraphics[width=6 cm]{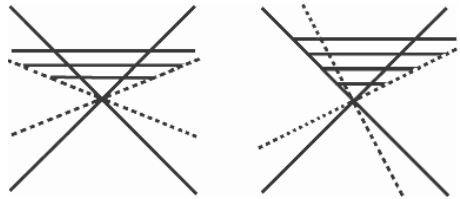}
\end{figure}

We shall rather define a extended notion of future-directed
timelike curves, by requiring that at each point of the curve, the
tangent vector is future-directed timelike with respect to
\textit{at least} one of the metrics $\mathbf{h}_i$. Such a
construction have been already advocated in \cite{MoffatClayton}.
All the (extended) notions of future, past, domains of dependence,
achronal sets, Cauchy surfaces, global hyperbolicity then follow.
This corresponds to a ``mixed'' notion of chronology: a point $P$
is in the (extended) future of $Q$ if it is in the future of $Q$
for at least one of the metrics $\mathbf{h}_i$, see
Fig.~(\ref{FigureFuture}). A spacetime interval is thus spacelike
in the extended sense if it is spacelike with respect to
\textit{all} metrics $\mathbf{h}_i$.

Causality then requires that a global (mixed) chronology on
spacetime exists. In other words, spacetime must be globally
hyperbolic in that extended sense. This straightforward
generalization enables us to express the miniminal formulation of
causality by the following properties:
\begin{enumerate}
\item[(a')] The mixed chronology defined by the set of metrics
$\mathbf{h}_i$ must be a global chronology on spacetime.
\item[(b')] The whole set of equations of motion must admit a
well-posed Cauchy problem.
\end{enumerate}
The formulation of causality (a)--(c) is immediately obtained from
these more general requirements whenever the matter light cones
coincide or lie within the gravitational one. We thus see that
fields can propagate superluminally without threatening causality
provided that we do not refer to any preferred chronology on
spacetime. The Cauchy problem of some field $\psi$ will be well
posed depending on the precise equation of motion, and if initial
data are set on surfaces that are spacelike with respect to the
metric $\mathbf{h}$ along which it propagates. Moreover, if we are
interested in the whole theory (gravity included), the Cauchy
problem will be well posed if initial data are set on surfaces
that are spacelike in the extended sense, that is
\textit{spacelike with respect to all metrics $\mathbf{h}_i$}).

\subsection{\label{CausaliteGlobal}On global properties}

It must be stressed that, in GR, the fact that $\mathbf{g}$
defines a global chronology cannot be proved, for the reason that
local physics does not determine the topology of spacetime, which
could however prevent the existence of a global chronology. This
the case for instance, of non time-orientable spacetimes.
Moreover, there exists exact solutions to the Einstein equations
that do not describe globally hyperbolic spacetimes, as explicitly
shown by the Kerr solution, which possess closed timelike curves,
or G\"odel's universe. Einstein equations thus admit solutions
that violate causality.

This difficulty is entirely subsumed into the so-called chronology
protection conjecture \cite{HawkingConjecture} that asserts that
the local laws of physics are such that they prevent the formation
of CFTCs in spacetime\footnote{At least CFTCs that are not hidden
behind an horizon.}. This has not been proved yet, so that we
usually assume from the beginning that spacetime
($\mathcal{M},\mathbf{g}$) is globally hyperbolic. As long as the
conjecture is not proved, such a restriction is not dynamical, but
of epistemological nature\cite{Bois}.

Let us stress that in our extended framework, and notably in
presence of superluminal matter fields, we will not be able to
prove either that the condition (a') holds. We will have to
\textit{assume} that spacetime is globally hyperbolic in the
extended sense. Again, if we were able to prove the chronology
protection conjecture, we would not have to make such a non
trivial assumption. We will discuss further this point in
Sec.~\ref{KessencesGlobal}.

In other words, the condition (a') will have to be imposed by
hand, just as in GR. As we already explained, we will not consider
the issue of the existence of the flow of time, and thus, in the
following, we will mostly be concerned by the point (b'). We
provide in the next section two examples of theories that do not
admit a well-posed Cauchy problem. It enables us to show the deep
relationship between non causal behaviors, closed curves in
spacetime and constraints on initial data.

\subsection{\label{Tachyons}Noncausal theories: closed curves and constraints on initial data}

An obviously non causal theory is an elliptic Klein-Gordon scalar
field with an equation of motion $h^{\mu\nu}
\partial_{\mu} \partial_{\nu} \varphi=0$, and where the metric $\mathbf{h}$
has the signature $\pm 4$. This elliptic equation does not have a
well posed Cauchy problem. It can be simply interpreted as the
fact that this Euclidean metric does not select the time
coordinate as special compared to the spatial one. There is thus
no available notion of propagation and initial data on
three-surfaces cannot be propagated in four dimensions. On the
contrary, a Lorentzian signature for $\mathbf{h}$ guarantees that
the equation is hyperbolic and that, by virtue of well-known
theorems \cite{Wald}, the Cauchy problem is well posed.

\begin{figure}
\caption{\label{FigureTachyons} The closed curve followed by the
tachyonic signal viewed in the rest frame of $A$. The tachyon is
sent by $A$ at time $t_0$ (event $E_0$) and received by $B$ at time $t_1$ (event $E_1$). Since
$B$ moves with respect to $A$, the tachyonic signal he sends back
to $A$ is actually received before it was sent, at time $t_2 <
t_0$ (event $E_2$). The thin line represents the Minkowski cone, and horizontal and vertical
lines are the space and time axis in the frame of $A$.}
\includegraphics[width=6 cm]{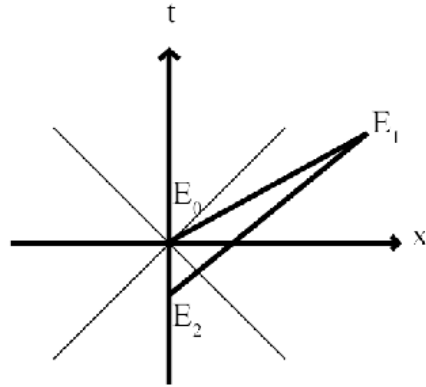}
\end{figure}

We consider now the more interesting case of tachyonic particles.
By tachyons we mean particles or signals that can be sent at a
superluminal speed relative to the emitter\footnote{Note that
``field theory tachyons'', i.e. negative mass-squared particles,
are \textit{not} of this type and are causal (see for instance
\cite{Susskind}, Appendix $B$)}. As correctly recognized in many
papers in the literature, tachyons are not causal (see, e.g.
\cite{antitelephone, LiberatiVisser}). Indeed, they can always be
used construct closed curves in spacetime along which a
(tachyonic) signal propagate. Consider for instance an observer
$A$ who sends at time $t_0$ a signal to observer $B$ (event
$E_0$), who in turn sends a signal back to $A$, at time $t_1$
(event $E_1$). This last signal can be received by $A$ at a time
$t_2 < t_0$ (event $E_2$), if signals can be sent from $B$ at a
superluminal speed and if $A$ and $B$ are in some relative motion
at a speed $v < c$, see Fig.~(\ref{FigureTachyons}). Information
can therefore propagate along the closed curve in spacetime
($E_0,E_1,E_2,E_0$).

The theory is not causal because an event along this curve will
influence itself, after the signal had gone one time round the
curve. Such an event is thus not freely specifiable, and we see
that propagation of information along a closed curve in spacetime
\textit{always} lead to constraints on initial data. On the
contrary, if the Cauchy problem were well posed for such a theory,
it would notably mean that there exists surfaces on which initial
data can be freely specified (up to possible constraints arising
from gauge invariance, if any) and unambiguously evolved with
time. It shows that the present theory cannot possess a well posed
Cauchy problem.

\subsection{\label{Paradoxes}Closed curves and temporal paradoxes}

Before going further, let us comment the previous experimental
device. Suppose (situation $1$) that, if $B$ receives a signal
from $A$, he sends a signal backwards in time to $A$. Then $A$
receives a signal at time $t_2$ and is thereby \textit{forced} to
send a signal to $B$ an instant later. It explicitly shows the
constraints that appear in the case of propagations along closed
curves.

A temporal (or causal) paradox can even arise if we consider the
following experimental device (situation $2$): when $B$ receives a
signal from $A$, he sends him back a signal, and $A$ sends a
signal to $B$ \textit{only if} he did not receive anything from
$B$. One may wonder what happens in that case. If $A$ receives a
signal from $B$, he will not send a signal to $B$ who in turn
would not send a signal back to $A$, and this contradicts the
hypothesis. The other possibility, when $A$ does not receive
anything from $B$, is also inconsistent. This is in fact a version
of the famous ``grandfather paradox''.

Remarkably, there is no answer to the above question ``what
happens?''. It is however immediate to see that this question
implicitly assumes that the knowledge of the initial data at time
$t_2$ for instance ($A$ receives a signal from $B$ or not) is
enough to determine what will happen ``later'' along the curve.
This question thus assumes that causality (or determinism) holds,
whereas it cannot even be defined, since there is no available
notion of time-ordering along a closed curve. On the contrary, we
have seen that such a closed curve leads to constraints on initial
data. The event at time $t_2$ is strongly correlated to itself so
that the only relevant question reads ``are any of the two initial
conditions: $A$ receives -or does not receive- a signal from $B$
allowed?''. Clearly the answer to this last question is negative.

Let us emphasize that situation $2$ (the grandfather's paradox) is
often taken as an example of bad causal behavior induced by
tachyonic particles (or accordingly, induced by the use of time
machines). This point is however not justifiable because the
paradox only arises if we ask the wrong question. This
experimental device simply \textit{cannot} exist since there is no
initial data consistent with it. There is nothing to do here with
causality, but only with logics: temporal paradoxes do not exist.
On the other hand, the noncausal situation $1$ can exist
(constrained initial data exists that correspond to this
experiment).

\subsection{\label{CausaTheoriesSuperluminiques}On superluminal behavior in field theories}

Tachyonic particles propagate superluminally and violate
causality. This does not mean however that any superluminal
propagations does. Indeed, the closed curve in
Fig.~(\ref{FigureTachyons}) can only be constructed if the speed
of the tachyonic signal depends on the speed of the
emitter\footnote{Explicitly, on assumes that the tachyon
propagates at some speed $u>c$ in the rest frame of $B$, and then
deduces, using a Lorentz boost, that it travels backward in time
in the frame of $A$, see \cite{LiberatiVisser}.}. This is not the
case in metric description of superluminal propagations that we
will mostly be concerned with.

Here we wish to investigate briefly what may look like a field
theory that involves superluminal propagation. In a theory whose
whole action is Lorentz-invariant, the only possibility to open up
the causal cones is to modify the propagational part of the
dispersion relation. It can be achieved by adding higher-order
derivatives in the action, but unless these terms arise from the
low energy limit of some UV-complete theory, the theory is
generically unstable \cite{Simon}. Superluminal behavior for
instance arises from quantum corrections to electrodynamics in non
trivial vacua, see Sec.~\ref{SectionQED}.

On the other hand, we could still have second order, but
nonlinear, equations of motions. We will consider two examples of
such theories in the following sections. This nonlinearity, which
is essential to obtain unusual dispersion relations, leads to a
background-dependent Cauchy problem. Accordingly, Lorentz
invariance of the action is spontaneously broken through non
trivial backgrounds and it allows superluminal propagations. Note
also that the speed of small perturbations around the background
only depends on the background and \textit{not} of the motion of
the emitter. Superluminal signals in field theories are not
tachyonic in the sense indicated in the previous section.

Note that we could also have considered non-Lorentz invariant
Lagrangians like $L=- \dot{\phi}^2 + u^2(\nabla \phi)^2$, where $u
\neq c$. We will not consider such Lagrangian in the present
paper. In practice, the theories we will consider will lead to
similar field equations, but only through a spontaneous breaking
of local Lorentz invariance by non trivial backgrounds.

\section{\label{SectionKessence}K-essence field theory}

\subsection{\label{KessenceBasique}Field equation and the Cauchy problem}
Let us consider a so-called k-essence scalar field theory, whose
action reads
\begin{equation}
\label{ActionKessenceMatiere} S=- m^4 \int \sqrt{-g} d^4x
\left[F\left(\frac{X}{m^2}\right) + V(\varphi)\right].
\end{equation}
where
\begin{equation}
\label{DefDeX} X \equiv \frac{g^{\mu\nu}\partial_{\mu}\varphi
\partial_{\nu} \varphi}{2},
\end{equation}
and $F$ and $V$ are some functions, $m$ some mass scale and
$\hbar=1$. Hereafter we also take $m=c=1$. Note that the more
general form $L=-F(X,\varphi)$ is often considered, but the
following arguments also apply to this case. Let us first neglect
the coupling to gravity which will be considered later, and
consider that $\mathbf{g}$ is flat. The field equation then reads
\begin{equation}
\label{EqChampKessenceMat} G^{\mu\nu}\partial_{\mu} \partial_{\nu}
\varphi -V'(\varphi)=0
\end{equation}
where the effective metric $G^{\mu\nu}$ is given by
\begin{equation}
\label{MetriqueG} G^{\mu\nu} \equiv F'(X) \eta^{\mu\nu} + F''(X)
\partial^{\mu} \varphi \partial^{\nu} \varphi,
\end{equation}
and a prime denotes derivation with respect to $X$. Scalar waves
propagate inside the ``scalar cone'' defined by this effective
metric. This cone generally differs from the gravitational one
unless $F''(X_0)=0$, see Sec.~\ref{SuperluminalKessence}.

This equation is a special case of quasilinear second order
differential equation. A theorem due to Leray then proves
\cite{Wald} that this equation has a well posed Cauchy problem if
spacetime ($\mathcal{M}, G^{\mu\nu}_0$) is globally hyperbolic,
where $G^{\mu\nu}_0$ is the effective metric evaluated on a
solution $\varphi_0$ to the field equation
Eq.~(\ref{EqChampKessenceMat}). Hereafter, we will refer to
$\varphi_0$ as the background. The initial data have to be
specified on three-surfaces that are spacelike with respect to the
background metric $G^{\mu\nu}_0$, in deep connection with the
discussion of Sec.~\ref{CausalityAndSuperluminalBehav}.

A necessary but not sufficient condition for global hyperbolicity
is the Lorentzian signature of the background metric
$G^{\mu\nu}_0$, which notably ensures that the field equation is
hyperbolic. By diagonalization of the matrix $G^{\mu\nu}_0$, we
find that its signature is $+2$ everywhere over $\mathcal{M}$ if
and only if
\begin{subequations}
\label{CondHyperKessenceMat}
\begin{equation}
\label{CondHyperPositive}
F'(X_0) > 0
\end{equation}
\begin{equation}
\label{CondHyperTime}
F'(X_0) + 2 X_0 F''(X_0)>0
\end{equation}
\end{subequations}
where $X_0 = \eta^{\mu\nu}\partial_{\mu} \varphi_0 \partial_{\nu}
\varphi_0 /2$. \textit{In particular}, if the function $F$ is such
that the above inequalities are satisfied for all $X$, the metric
$G^{\mu\nu}_0$ will have a signature $+2$ on any background.
Notice that the sign of the above inequalities
Eq.~(\ref{CondHyperKessenceMat}) can be reversed. This would
however correspond to a metric $G^{\mu\nu}_0$ with signature $-2$
and the scalar waves would carry negative energy. The theory would
therefore be unstable when coupled to other fields (notably
gravity). We exclude here this possibility, although it should be
stressed that it comes from a stability argument, and not from
some causal requirement.

In presence of gravity the Cauchy problem has to be solved
simultaneously for the gravitational variables and for the matter
ones. It has been proved (see for instance \cite{Hawking}) that
the Cauchy problem is well posed if matter fields satisfy
``reasonable'' equations of motion\footnote{Notably when they form
a quasilinear, diagonal and second order hyperbolic system of
equations.} and if the stress-energy tensor of the matter fields
only involve matter variables, the gravitational metric, and their
first derivatives. This is notably the case of k-essence theories,
as can be easily checked. Locally therefore, the whole theory has
a well-posed Cauchy problem.

\subsection{\label{SuperluminalKessence}Superluminal behavior}

The analysis of the characteristic\footnote{See
\cite{CourantHilbert}, Ch. 5, Appendix 1.} of the field equation
Eq.~(\ref{EqChampKessenceMat}) shows that the scalar field
propagates superluminally if $F''(X_0) > 0$, where we used that
$F'(X_0) > 0$ by virtue of Eq.~(\ref{CondHyperPositive}).

As it will be useful later, let us remind that any field which
propagates along an effective metric of the form (up to some
positive conformal factor):
\begin{equation}
\label{MetriqueEffective} H^{\mu\nu}= g^{\mu \nu} + B
n^{\mu}n^{\nu},
\end{equation}
where $B$ and $n^{\mu}$ are respectively some scalar and vector
field, is superluminal if $B>0$. Notice that the opposite sign for
$B$ is found in \cite{Susskind,Arkani-Hamed} because of the
opposite choice of signature.

\subsection{\label{CauchySurfArkaniSusskind}``Local causality'' and the choice of initial data surfaces}

The above theorem and our analysis of the meaning of causality in
Sec.~\ref{Causalite} enable us to conclude that k-essences
theories are causal in flat spacetime whenever the background
($\mathcal{M}, G_0^{\mu\nu}$) is globally hyperbolic, even in
presence of superluminal scalar waves. Again, this conclusion
holds only if we do not refer to any preferred chronology.

It was however claimed \cite{Susskind,Arkani-Hamed,
BekensteinMilgrom84,Bekenstein2004} that k-essences theories are
not causal even if the background is globally hyperbolic. Here we
emphasize that these claims can only be supported by the postulate
that the gravitational (here, the flat) metric induces a preferred
chronology on $\mathcal{M}$, a point that, we believe, is
implicitly assumed in these references. The main argument which
rules out superluminal propagations is indeed that the Cauchy
problem for the scalar field is not well posed for initial data
that are set on surfaces which are spacelike \textit{with respect
to the flat metric} but timelike or null with respect to the
background metric. In that case, initial data cannot be evolved
because of caustics \cite{Arkani-Hamed}, and the Hamiltonian
formalism is singular \cite{Susskind}.

Let us however stress that it is not surprising in view of Leray's
theorem, since it proves that initial data surfaces for the scalar
field must be spacelike \textit{with respect to the background
metric}. If initial data are set on these surfaces, the theory is
free of such singular behaviors.

This claim therefore only arises from an unadapted choice of
initial data surfaces, i.e., from the postulate that the
gravitational metric defines a preferred chronology. This
postulate actually prevents \textit{any} superluminal propagation,
as we explained in Sec.~\ref{CausaliteGR}. In such a case,
superluminal behavior is ruled out \textit{from the beginning},
but not by some intrinsic (mathematical) arguments.

\subsection{\label{KessencesGlobal}Global properties}

The only threat for causality lies in fact in global properties of
the background. It has been correctly recognized in
\cite{Arkani-Hamed} that the spacetime ($\mathcal{M},
G^{\mu\nu}_0$) may not be always globally hyperbolic. In
particular closed timelike curves (with respect to the effective
metric) could exist.

Let us first stress that in general however, the spacetime
($\mathcal{M}, G^{\mu\nu}_0$) is globally hyperbolic. It is
notably the case in trivial backgrounds $\varphi_0 =
\textrm{const.}$, and in non trivial but homogeneous (in a certain
Lorentz frame) backgrounds $\partial \varphi_0 = \textrm{const.}
\neq 0$, which may be relevant in a cosmological context. Note
that these backgrounds have to be solutions to the equation of
motion Eq.~(\ref{EqChampKessenceMat}). In the first case,
$\varphi_0$ has thus to be an extremum of the potential, whereas
in the second one, $\varphi_0$ is not constant in spacetime and
the potential must be flat over some range, or more simply vanish.
In these two important cases, the spacetime ($\mathcal{M},
G_0^{\mu\nu}$) is globally hyperbolic and, by virtue of the
previous theorem, the Cauchy problem is well posed and the theory
is causal.

Since the global hyperbolicity of spacetime ($\mathcal{M},
\mathbf{g}$) must be assumed in GR, as we explained in
Sec.~\ref{CausaliteGlobal}, we may as well assume that spacetime
is globally hyperbolic in the extended sense, see
Sec.~\ref{CausalityAndSuperluminalBehav}. Such an assumption
obviously requires that Eq.~(\ref{CondHyperKessenceMat}) hold, and
automatically ensure that the whole theory of the k-essence scalar
field and gravity is causal. The relevant Cauchy surfaces in that
case are hypersurfaces that are spacelike with respect to all
metrics (gravitational, scalar and others matter metrics).

Note that this is a non trivial (and maybe arguable) assumption.
Let us however stress that it must also be assumed in the context
of GR and standard matter alone so that the fact that non trivial
global properties may break causality does not appear to be rooted
in superluminal propagations. This assumption may actually be
linked to the chronology protection conjecture, as shown by the
following example. Let us consider the highly non trivial and non
globally hyperbolic background invoked in \cite{Arkani-Hamed}. It
consists of two ``bubbles'' of non trivial backgrounds $\partial
\varphi_0 = \textrm{const.} \neq 0$ that move rapidly in opposite
directions with a finite impact parameter. The space is otherwise
empty (trivial background $\varphi_0 = \textrm{const.}$). Small
scalar perturbations thus travel superluminally inside the two
bubbles, and along null rays of the flat metric outside them. The
fact that the two bubbles are in relative motion implies the
existence of CFTC (timelike with respect to the background metric
$G^{\mu\nu}_0$).

It should however be stressed that such a background is
\textit{not} relevant since it is not a solution to the equation
of motion Eq.~(\ref{EqChampKessenceMat}). Indeed, the derivative
of the field $\varphi_0$ is not continuous at the transition
between the bubbles and the empty space. Some Dirac-like source
terms should be added to the equation of motion to accommodate
such a background. On the contrary, a physical solution involving
such two bubbles should exhibit a continuous transition from their
interior to the empty space. Of course it may turns out that even
in this more realistic situation the background is still not
globally hyperbolic, but the contrary could also happen, and it
would be an illustration of the chronology protection conjecture.
We did not try to perform this analysis, but it may be
interesting.

A very similar case was discussed in \cite{LiberatiVisser} in the
context of Casimir experiment, see Sec.~\ref{SectionQED}. Note
that an analogous case has also been found in the context of GR by
Gott \cite{Gott}, who showed that two straight infinite cosmic
strings moving in opposite directions with a finite impact
parameter lead to the formation of CFTC in spacetime. Again, it
shows that these difficulties with causality at a global level
already exists in the context of GR and subluminal matter, and
should not be related to some intrinsic problems of superluminal
propagations.

\section{\label{Bimetrictheories}Bimetric theories of gravity}

\subsection{\label{BimetricDef}Definition}
By multi-metric theories of gravity we mean theories of gravity in
which some degrees of freedom of the matter sector are coupled to
some matter metrics $\tilde{\mathbf{g}}_i$ distinct from the
gravitational one $\mathbf{g}$. The fact that different matter
fields are coupled to different metrics breaks the weak
equivalence principle (WEP), which has been tested with great
accuracy. As a special case, a bimetric theory of gravity is a
theory where all the matter fields are coupled to the same metric
$\tilde{\mathbf{g}}$. This ensures that the WEP is satisfied. GR
is just the special case $\mathbf{g}=\tilde{\mathbf{g}}$.

A typical example is scalar-tensor theories of gravity in which
$\tilde{\mathbf{g}}= \Omega^2 \mathbf{g}$, at each point of
$\mathcal{M}$, and where $\Omega$ is a smooth real-valued function
over $\mathcal{M}$. Dynamics of the scalar field $\Omega$ arises
from a standard kinetic term (but could also be of k-essence
type). Because of the conformal relationship between $\mathbf{g}$
and $\tilde{\mathbf{g}}$, the two cones defined by these two
metrics coincide, and there is no superluminal propagations.

In general however, we could have non conformal relation between
these two metrics. Consider for instance, the so-called disformal
relation \cite{BekensteinDisformal}
\begin{equation}
\label{DisformalMetric} \tilde{g}_{\mu\nu}=A^2 \left(g_{\mu\nu} +
B U_{\mu} U_{\nu}\right),
\end{equation}
where $A$ and $B$ are some functions of the scalar quantity
$U_{\mu}U^{\mu}$, and $U$ is a vector field. When this matter
metric is of Lorentzian signature, the matter light-cone (defined
by $\tilde{\mathbf{g}}$) can be wider than the gravitational cone
(defined by $\mathbf{g}$), depending on the sign of $B$. This
disformal relation is an essential piece of some recent
relativistic field theories of the MOND paradigm
\cite{Bekenstein2004, Sanders97}.

\subsection{\label{BimetricSuperlum}Superluminal behavior}

Throughout this paper, we defined superluminality as the
propagation along spacelike curves of the gravitational metric.
But we could also have defined it as going faster than light
(photons). These definitions coincide in GR but not in bimetric
theories since $\mathbf{g} \neq \tilde{\mathbf{g}}$. In that
framework, there are two opposite definitions of superluminal
behaviors. Photons can travel faster than gravitational waves and
conversely. It is thus not clear which of these two superluminal
behavior we have to worry about.

This is closely related, again, to the choice of a preferred
chronology. If one assumes that the gravitational metric field
induces a preferred chronology with respect to which causality
should be defined, then, as in Sec.~\ref{CausaliteGR}, the matter
light-cone must coincide or lie within the gravitational one
everywhere on spacetime. It was, on the contrary, claimed in
\cite{BekensteinDisformal,Bekenstein2004} that, since rods and
clocks are made of matter, the matter metric $\tilde{\mathbf{g}}$
should be somewhat favored, in the sense that in order to preserve
causality, no propagations should escape the matter light-cone. In
particular, this cone should be wider than the gravitational one.
Very interesting is the fact that this postulate is the exact
opposite of the previous one. It is perhaps the best way to show
that there is no clear reason why one metric should be preferred
to the other.

This confusion can be easily understood. Let us go further in the
discussion of Sec.~\ref{CausalityAndSuperluminalBehav}. Whenever
$\tilde{\mathbf{g}}$ is Lorentzian and disformally related to
$\mathbf{g}$, we have at hand two metrics which reduce locally to
constant metrics $\mathbf{g}_0$ and $\tilde{\mathbf{g}}_0$. As a
consequence, there exist two classes of inertial coordinates for
which one of these two metrics \textit{but not both}, reduce to
its fundamental form $\boldsymbol{\eta}$. Inertial coordinates of
each class transform under the action of the Lorentz group
$SO(3,1)$ with a \textit{different invariant speed}\footnote{In a
sense, our analysis completes the interesting discussion about the
different facets of $c$, see \cite{UzanEllis}.}.

Let us emphasize that in GR, superluminality is defined as the
propagation on spacelike curves with respect to the gravitational
metric mainly because the latter is thought to be ``the''
spacetime metric, since it reduces locally to $\boldsymbol{\eta}$.
We already stressed in Sec.~\ref{CausalityAndSuperluminalBehav}
that any Lorentzian can actually take this form locally, so that
it could also be viewed as the spacetime metric. The framework of
bimetric theories considerably enlighten this point. In a sense,
indeed, we have two ``natural'' metrics and there is simply no way
to decide which of these two metrics should be ``the'' spacetime
metric. Accordingly, there are two natural chronologies and two
locally invariant speeds. The above two classes of inertial
coordinates corresponds to rods and clocks made of matter or
gravitons.

Bimetric theories thus clearly show the irrelevance of postulating
a preferred chronology. Let us also stress that without such a
postulate, the notion of superluminal behavior becomes itself
irrelevant. The only point, as far as bimetric theories are
concerned, is actually that the ratio of the speed of
gravitational waves to the speed of photons for instance, vary in
space and time. This is the reason why such a framework should be
considered as the best motivated one that reproduces Varying Speed
of Light theories, see Sec.~\ref{BimetricVSL}.

\subsection{\label{BimetricCausalite}The Cauchy problem and global properties}

Following the discussion of
Sec.~\ref{CausalityAndSuperluminalBehav}, let us define causality
by (a')--(c'). It is clear that the matter metric has to be
Lorentzian for the Cauchy problem to be well posed. This condition
reads $1 + B U_{\mu} U^{\mu}> 0$ and $A$ must be non zero. We
wrote the conformal factor as $A^2$ in order to ensure that matter
fields carry positive energy. The matter stress-energy tensor only
depends on the matter fields, the vector field, the gravitational
metric, and their first derivatives. Depending on the precise form
of the action of the vector and matter fields, the whole set of
equations of motion (including gravity) may be a diagonal, second
order, and quasilinear hyperbolic system. Note that when $U$ is
given by the gradient of a scalar field $U_{\mu} = \nabla_{\mu}
\varphi$, the initial value formulation of the scalar equation is
a quite involved question, see Sec.~\ref{BimetricScalaire}. On the
other hand, when $U$ is a ``true'' vector field, the Cauchy
problem is generically well posed\footnote{If the kinetic term of
the vector field differs from the usual (Einstein-Maxwell) one,
the vector field equation involves second derivatives of the
metric field $\mathbf{g}$ and is thereby not diagonal. It does not
mean however that the Cauchy problem is not well posed, but rather
than a careful analysis is in order. Generic vector field actions
have been considered in \cite{Jacobson} (and references
therein).}.

We have also to consider the global structure of spacetime. As in
the case of GR, or GR and k-essences, we have to assume that
spacetime is globally hyperbolic in the extended sense of
Sec.~\ref{CausalityAndSuperluminalBehav}. This assumption,
together with the above conditions on equations of motion of the
matter fields and the vector field ensures that the whole theory
is causal.

\section{\label{SectionQED}Quantum induced superluminal propagations}

In quantum electrodynamics, effects of temperature,
electromagnetic or gravitational backgrounds, or boundaries (e.g.
Casimir plates) break local Lorentz invariance at the loop level
through vacuum polarization, thus leading, in some cases, to
faster-than-$c$ light propagation \cite{FullList, Shore,
LiberatiVisser}; see also \cite{Scharnhorst} and references
therein. Note that these results are derived within some range of
approximation using the effective action formalism, up to the one
or two loop level. Generically the results only hold at
frequencies less than the electron mass $m_e$. One therefore only
derives the phase velocity of soft photons, whereas the actually
relevant speed is the wavefront velocity, which is the phase
velocity at infinite frequency and which corresponds to the
analysis of the characteristics \cite{Shore}.

The computation of wavefront velocities is a non-perturbative task
that we will not be concerned with. An argument based on the
standard Kramers-Kronig relation \cite{Scharnhorst} actually gives
some hints on the value of the wavefront velocity. In the case of
Casimir vacua, it was shown that the wavefront velocity might be
greater than $c$ in the direction orthogonal to the plates
(breaking of Lorentz invariance by the boundaries). The wavefront
velocity has to be equal to $c$ in the parallel direction because
Lorentz invariance is left unbroken in that direction, at least if
the Casimir plates are infinite (or if boundary effects are
negligible).

Let us assume the validity of this result. Of course, ``light does
not travel faster than light'', but the point is that gravity does
not see the plates in first approximation, so that Lorentz
invariance is not broken in the gravitational sector and gravitons
still propagate at the velocity $c$. The ratio of the speed of
photons to the one of gravitons is thus greater than one.

Our discussion of Sec.~\ref{Causalite} enables to give an
immediate answer to the question of causality in that case. It can
be shown that photons propagate inside the plates along an
effective metric of the form of Eq.~(\ref{MetriqueEffective}),
where $B$ is some positive constant in that case, and $n^{\mu}$ is
the unit spacelike vector orthogonal to the plates (see
\cite{LiberatiVisser} and references therein). Outside the plates,
photons propagate along the flat metric $\boldsymbol{\eta}$ (here
we neglect curvature).

It was correctly recognized in \cite{LiberatiVisser} that such a
metric is stably causal, so that photons \textit{cannot} propagate
along closed curves, contrary to the claim made in
\cite{DolgovAdetruire} (see also the criticism of
\cite{DolgovDetruit}). Actually, this spacetime is even globally
hyperbolic and therefore \textit{perfectly causal}, even if
photons propagate faster-than-$c$ inside the plates.

Causality may however be lost if two Casimir vacua are moving
rapidly towards each other \cite{LiberatiVisser}. In that case,
spacetime may possess CFTC (with respect to the effective metric).
This case is strictly analogous to the case of two bubbles made of
non trivial background in k-essence theory, see
Sec.~\ref{KessencesGlobal}. Authors of \cite{LiberatiVisser} then
invoked the chronology protection conjecture, and notably noted
that the two Casimir vacua should be confined within plates that
cannot be infinite, so that non trivial boundaries effects may
prevent the formation of CFTC. We essentially reached the same
conclusion in Sec.~\ref{KessencesGlobal}. Similar arguments may be
applied to others vacua. In particular, note that vacuum
polarization induces some effective metric along which photons
propagate. This effective metric only differs from the flat one by
terms of order $\alpha^2$ where $\alpha$ is the fine structure
constant. These corrections are therefore small (at least in
``reasonable'' vacua), so that the effective metric is still
Lorentzian and the spacetime is still globally hyperbolic.

\section{\label{Applications}Applications}

We briefly comment some applications of our results to recent
interesting developments on possible modification of gravity.

\subsection{\label{DEetGhosts}K-essences theories and dark energy}

K-essence scalar field theories have been suggested as promising
candidates of the dark energy fluid \cite{Chiba} (see also the
review \cite{ReviewDE} and relevant references therein). Such an
effective action can also be motivated by the low energy regime of
some string theories \cite{Damour}. The hyperbolicity conditions
Eq.~(\ref{CondHyperKessenceMat}) on the function $F$ have been
correctly derived in the literature, but in a different way
(except in \cite{Haloesofkessence}).

Authors usually require the stability of scalar perturbations
around some backgrounds. It depends on the reality of the ``sound
speed''
\begin{equation}
c_s^2=\frac{F'(X)}{F'(X)+2 X F''(X)},
\end{equation}
and is thus equivalent to the Lorentzian character of the
effective metric $G^{\mu\nu}$. This (un)stability is therefore
directly related to the hyperbolic (resp. elliptic) character of
the scalar field equation. Note that the signature can still be
$+2$ or $-2$. Authors then demand that these small perturbations
carry positive energy, and it implies $F'(X) \geq 0$. Here we wish
to stress that these results also arise from the analysis of
causality, and, moreover, that they are actually non-perturbative
results. It can be shown, indeed, that the positivity of the whole
Hamiltonian and not only the one of perturbations, is guaranteed
whenever the two conditions Eq.~(\ref{CondHyperKessenceMat}) hold
\cite{nous}.

Unusual kinetic terms have also been considerably debated because
phantom scalar fields can reproduce an equation of state $w < -1$
\cite{Caldwell, Chiba, ReviewDE}, which indeed reads
\begin{equation}
w=\frac{-F(X)}{F(X)-2 X F'(X)}.
\end{equation}
Whenever the density $\rho = F(X)-2 X F'(X)$ is positive $w<-1$ is
equivalent to $F'(X) < 0$. Thus only phantom (ghost) matter can
lead to super-acceleration of the Universe. Since ghosts are
usually associated with a fatal instability at the quantum level
and notably in the UV regime, authors have suggested a
stabilization mechanism of the ghost field at the UV scale. The
function $F$ could be such that $F \sim -X/m^2
+\mathcal{O}(X^2/m^4)$ (we reestablish the mass scale $m$). In
that case the ghost only appear at low energy $X \ll m^2$, but
could be stabilized at higher energies by higher-order terms, and
the timescale of the instability can be made arbitrarily high
\cite{Carrollghost, Cline}. Such a mechanism may however suffers
serious diseases. Notice indeed that the sign of $F'(X)$ must
change for some values of $X_c$ and the squared sound speed may
become negative. Accordingly, the effective metric may become
Euclidean and the Cauchy problem will not well posed anymore. The
hyperbolicity seems however still guaranteed if $F'(X) + 2 X
F''(X)$ also changes of sign at the same value $X_c$. Notice that
the function $F$ must then be quite fine-tuned : $X_c$ may be
equal to $0$ or $X_c$ must be a turning point as well as an
inflexion point of $F$. Moreover, in that already fine tuned case,
the effective metric becomes totally degenerate (vanishes) at the
point $X_c$. There is a caustic, and the theory is not well
defined.

It has been shown in \cite{GhostcondensatesAH} that this point
$X_c$ is however not reached through the cosmic evolution (or at a
time $t=\infty$), and the theory may thus be free of singular
behaviors. However the cosmological background is not the only
relevant one. The theory must also apply at local (e.g.
astrophysical) scales, and we expect the scalar field to have some
inhomogeneities. Local physics then drives $X$ to positive values,
and the above singular point $X_c$ will generically be crossed.
Moreover, at a quantum level, it is not clear if we can still make
sense of summing momentum from zero to some cutoff, whereas the
propagator is not defined for some value of the momentum.

In conclusion, k-essences field theories are quite relevant causal
theories that can account for the dynamics of the dark energy
fluid with an equation of state $w>-1$. On the other hand,
k-essences phantom theories, even if stabilized, suffer serious
diseases and it is very unlikely that such a fluid could drive the
super-acceleration of the Universe $w<-1$, if any. Note that it
was recently found that inhomogeneities of the matter distribution
in the Universe may be described by an effective scalar field
which could be a phantom in certain cases \cite{Larena}. Because
only effective, this field does not suffer from quantum
instabilities.

\subsection{\label{RaqualEpsilon}K-essences theories and the MOND paradigm}

K-essences theories are also used to account for the mass
discrepancy in galaxies and clusters, without the need for dark
matter. Milgrom \cite{Milgrom83} first pointed out that rotation
curves of spiral galaxies exhibit a discrepancy at an universal
acceleration scale $a_0 \sim 1.2 \times 10^{-10}
\textrm{m.s}^{-2}$, and therefore that a modification of Newton
dynamics (MOND) of the form
\begin{equation}
a \mu\left(\frac{a}{a_0}\right)= g,
\end{equation}
where $g$ is the Newtonian gravitational field and $a$ the
acceleration, could account for the observed discrepancy without
dark matter. The $\mu$ function must behave asymptotically as
$\mu(x)=1$ if $x \gg 1$ and $\mu(x)=x$ if $x \ll 1$. The $\mu$
function is otherwise free. The choice $\mu(x)=x/\sqrt{1+x^2}$ is
a quite standard one which fits well the data. Such a behavior in
the low acceleration regime automatically leads to flat rotation
curves far from the source, and also reproduces the well
established Tully-Fisher law $v^4 \propto L$, where $v$ is the
plateau velocity and $L$ is the luminosity of the galaxy.

This successful phenomenology \cite{Sanders2002} can be reproduced
with the help of relativistic aquadratic Lagrangians (RAQUAL)
\cite{BekensteinMilgrom84}, i.e., a k-essence scalar field.
Consider indeed a scalar field $\varphi$ coupled to matter via a
conformal metric $\tilde{\mathbf{g}}= \textrm{exp}(-\alpha
\varphi) \mathbf{g}$. The scalar field equation then reads
\begin{equation}
\nabla_{\mu} \left( F'(X) \nabla^{\mu} \varphi \right)= - 4 \pi G
\alpha T,
\end{equation}
where $\nabla$ is the covariant derivative corresponding to the
metric $\mathbf{g}$, $G$ is Newton's constant appearing in the
Eintein-Hilbert action, and $T$ is the trace of the stress-energy
tensor of the matter fields (defined by variation and contraction
with respect to $\mathbf{g}$). If we consider a static
distribution of matter $\rho$, this equation reduces to a static modified
Poisson equation for the scalar gravitational potential $\varphi$:
\begin{equation}
\nabla(F'\left((\nabla \varphi)^2 \right) \nabla \varphi) = 4 \pi
G \alpha \rho.
\end{equation}
It is then clear, after restoring the appropriate dimensionfull
constants, that the function $F'(x^2)$ plays the role of the
Milgrom's function $\mu(x)$. Note that $X>0$ in the static case.
If we require that $F(X)$ behaves as
$X$ if $X \gg 1$, and as $2/3 X^{3/2}$ if $X \ll 1$, we thus
recover the MOND phenomenology.

Note that, since the $\mu$ function is generally taken as a
monotonic (increasing) function of its argument, the second
derivative of $F$ is positive and the scalar waves propagate
superluminally. The theory were thus thought to be acausal
\cite{BekensteinMilgrom84, Bekenstein2004}. We however argued in
Sec.~\ref{Causalite} and Sec.~\ref{SectionKessence} that this
conclusion is correct only if the gravitational metric defines a
favored chronology, an assumption that may be dropped.

A critical point is however that the free function $F$ (related to
$\mu$) must satisfy the conditions
Eq.~(\ref{CondHyperKessenceMat}). It is however immediate to note
that the asymptotic conditions on $\mu$ implies that when $X$ goes
to $0$, $F'(X)$ and $F'(X) + 2 X F''(X)$ also go to zero and the
effective metric is completely degenerate.

What it means is that, in such a theory, in an astrophysical
context, there must exist around each galaxy or cluster a singular
surface on which the scalar degree of freedom does not propagate.
The reason is that, near the source, $X$ must be positive, but
negative far from it due to the cosmological background. This
theory can therefore not lead to a consistent picture of local
physics imbedded into a cosmological background. This major
objection also applies to the recent relativistic model of MOND of
\cite{Bekenstein2004}.

A trivial modification of the asymptotic form of the $\mu$
function however cures the problem. Let us consider that $\mu(x)
\sim x + \varepsilon$ if $x \ll 1$, or equivalently that $F'(X)
\sim \sqrt{X} + \varepsilon$ if $X \ll 1$. It ensures that the
theory is well behaved at the transition between local and
cosmological physics. This slight modification induces a
significant change in the phenomenology because there is a return
to a Newtonian behavior very far from the source, with a
renormalized value of the gravitational constant. Such a theory thus predicts that
rotation curves are only approximatively flat on a finite range
of $r$, and current data requires $\varepsilon$ to be at most of order
$1/10$. More details on this can be found in \cite{nous}.
Interestingly, in that kind of theory, MOND only appears
as an intermediate regime between two Newtonian ones only
differing by the value of the gravitational constant, the
transition, driven by the scalar field, occurring at Milgrom's
acceleration scale.

\subsection{\label{BimetricScalaire}Bimetric theories and MOND}

MOND-like theories of gravity must also predict enhanced light
deflection in order to be consistent with the data. In the
previous theory however, the conformal coupling of the scalar
field to the matter metric implies that light is not coupled to
the scalar field because of the conformal invariance of
electromagnetism in four dimensions.

This led authors to consider more general bimetric theory of
gravity in which matter is coupled to a disformal metric of the
type Eq.~(\ref{DisformalMetric}) \cite{BekensteinDisformal,
BekensteinDisformalApplication}. In these first models, the vector
field $U$ in Eq.~(\ref{DisformalMetric}) was assumed to be the
gradient of the k-essence scalar field, and $A$ and $B$ were some
functions of $\varphi$ and $X$. It was however proved that when
the matter light-cone is wider than the gravitational one (see
Sec.~\ref{Bimetrictheories}) then there is actually less light
deflection than in GR.

We have seen however that causality does not require the matter
light-cone to be wider than the gravitational one. On the
contrary, it could be inside the gravitational one and the light
deflection would be enhanced (compared to GR). Such a framework is
therefore relevant for relativistic theories of the MOND paradigm,
as we stress in \cite{nous}. The Cauchy problem is very likely to
be well posed, since the (Einstein) equation of the metric field
is still diagonalized, hyperbolic and of the second order. The
scalar field equation in vacuum has a well-posed Cauchy problem.
The scalar field equation inside the matter is however quite
complicated and is not diagonalized. The Cauchy problem inside the
matter is thus a quite involved question, but we expect that some
generic requirements on the free functions $A$ and $B$ and on
energy conditions of the matter sector will guarantee it, and by
the way, that the whole theory has a well posed Cauchy problem.
This is left for further investigation (see also \cite{nous}).

\subsection{\label{BimetricPioneer}Bimetric theory of gravity and the Pioneer anomaly}

Let us briefly consider another application of bimetric theories
of gravity. Conventional physics did not succeeded so far in
explaining the anomalous motion of the two Pioneer spacecrafts
\cite{Turyshev} that experience a small anomalous acceleration
(roughly directed towards the Sun).

Unconventional theories of gravity could however explain it. The
two Pioneer spacecrafts are moving along geodesics of the matter
metric $\tilde{\mathbf{g}}$, and we shall look for a slight
modification of the Schwarzschild solution to explain the Pioneer
anomaly. It is worth noting that this anomaly was detected soon
after Jupiter's flyby \cite{Turyshev} so that we have to check
that the suggested modification is compatible with the motion of
outer planets. Outer planets are moving along quasi-circular
orbits, and are thereby mostly sensitive to the time-time
component of the matter metric $\tilde{g}_{00}$. Very sensitive
tests of Kepler's third law \cite{ContraintesKepler} then put
stringent bounds on the deviation from the leading order of the
time-time component of the Schwarzschild metric. These bounds are
actually too small to account for the Pioneer anomaly.

This fact leads to the wrong statement in \cite{Pionnul} that the
Pioneer anomaly cannot be of gravitational origin. This is indeed
not justifiable since the two spacecrafts evolve on hyperbolic
trajectories and are thereby also sensitive to the radial-radial
component of the matter metric, which is not well constrained in
the outer solar system. Let us stress that, if spherical symmetry
is assumed, a slight modification of $\tilde{g}_{rr}$ compared to
the Schwarzschild solution is the only way to give the Pioneer
anomaly a gravitational origin in a metric theory of gravity. This
was notably realized in \cite{Reynaud}.

Here we provide, with the help of bimetric theories of gravity, a
framework that realizes this modification of the radial-radial
component. The action of the theory is given by the
Eintein-Hilbert action, a canonical action for the scalar field,
and the matter fields are coupled to the matter metric
\begin{equation}
\label{MetriqueDisfScalaire} \tilde{g}_{\mu\nu}=A^2(\varphi)
\left(g_{\mu\nu} + B(X) \nabla_{\mu} \varphi \nabla_{\nu} \varphi
\right),
\end{equation}
where $X$ is still defined by Eq.~(\ref{DefDeX}). As we already
stressed, non trivial properties of $A$, $B$ and energy conditions
of the matter sector may be required to ensure the hyperbolicity
of the scalar equation inside matter. In addition, the matter
metric has to be Lorentzian and this reads $1+ 2 X B(X) > 0$. In a
static and spherical symmetric situation, we have
$\tilde{g}_{00}=A^2 g_{00}$ and $\tilde{g}_{rr}=A (g_{rr}+ B(X)
\nabla_r \varphi \nabla_r \varphi)$. The bare metric $\mathbf{g}$
is given by the solution of Einstein equations and it can be shown
that it coincides with the Schwarzschild solution to the leading
order. This theory is thus a realization of the above
phenomenology. Such models predict ``Pioneer-like'' anomalies in
the precession of perihelion that could notably be found in
precise measures of the orbit of Mars. More details on the Pioneer
anomaly and disformal theories can be found in \cite{nous}.

\subsection{\label{BimetricVSL}Varying Speed of Light theories}

The first varying speed of light theories were constructed by
replacing $c$ by $c(t)$ in the equations of motion of GR, where
$t$ could be the cosmic time \cite{VSLBarrowetal}. This however
leads to equations of motion that do not conserve stress-energy
anymore. In other words, this theory cannot be obtained from a
variational principle \cite{VSLVisser,UzanEllis}.

Some authors thus made use of the disformal matter metric
Eq.~(\ref{MetriqueDisfScalaire}) to reproduce VSL within a
consistent framework
\cite{VSLMoffat,VSLVisser,InflationDisformal}. Note that in all of
these models, the free functions were taken as constants $A=1$ and
$B=-1/m^{2}$. Let us stress that if one insists on the Weak
Equivalence Principle there cannot be any coupling between
$\varphi$ and the standard model matter other than through the
matter metric. Then, by varying the action with respect to
$\varphi$, one finds that the scalar field is not created by the
matter sources ($\nabla_{\mu} \varphi =0$ is always a solution).
The scalar field can only be generated by some (non constant)
function $A(\varphi)$, like in scalar-tensor theories.

Note also that the choice of $B=-1/m^2$ in these works does not
guarantee the Lorentzian character of the matter metric, which can
actually be Euclidean in a cosmological background such that
$(\partial_0 \varphi)^2 > m^2$. One could however circumvent this
problem by arguing that the theory is only an effective model
valid for $(\partial \varphi)^2 \ll m^2$, and where $m$ could be
of the order of the GUT's scale.

As we already stressed, the initial value formulation of the
scalar field equation in such models is a quite involved question.
This point has not been pointed out although the existence of an
initial value formulation is a basic requirement that must be
checked in order to give any sense to such VSL models.

Let us finally remark that, since the fine structure constant
$\alpha$ is proportional to the inverse of the speed of light, it
has been argued that VSL theories could also account for the
variation of $\alpha$ with cosmic time, if any \cite{Webb,petitjean}.
However, in this bimetric framework of VSL theories, if one
analyzes atomic rays at some redshift by usual techniques, one is
observing electromagnetic phenomenon using matter rods and clocks,
so that no variations of $\alpha$ are actually observable. The
fact that the ratio of the speed of light to the speed of
gravitational waves varies in space and time does not lead to a
variation of $\alpha$ (if it is measured in the usual way), but
simply to a redefinition of the redshift $z$ of distant objects.

\section{\label{Conclusions}Conclusions}

We provided a careful analysis of the meaning of causality in
classical field theories. This led us to the conclusion that
superluminal behaviors are generically found to be non causal
\textit{only if} one refers to a \textit{prior} chronology on
spacetime. This postulate actually states that, locally, some sets
of inertial coordinates must be preferred ones, and this appears
to be in great conflict with the spirit of general relativity, and
more precisely with the spirit of general covariance. On the
contrary, we derived in Sec.~\ref{CausalityAndSuperluminalBehav},
by means of the conditions (a')--(c'), a formulation of causality
in which coordinates are still physically meaningless.

Note that while referring to a preferred chronology may seem
natural in GR since all fields (both gravity and matter) propagate
along the gravitational metric $\mathbf{g}$, it becomes somewhat
unnatural whenever spontaneous breaking of Lorentz invariance
occur, be it driven by quantum polarization (and in general, the solution of field equations),
nonlinearities of some new fields, etc. The resulting spacetime is generically
endowed with a finite set of Lorentzian metrics $\mathbf{h}_i$
which may not be conformally related to each others. In that case
rods and clocks made of different fields lead to different systems
of coordinate that do not transform under the same Lorentz group.
There is not only one, but at least two invariant speeds in that
case. Note that it would be misleading to think that there is a preferred
Lorentz invariance in the theory because of the symmetry of the action.
Actually, in a GR-like context, the action is not Lorentz invariant but
diffeomorphism invariant. This notably means that all local system of
coordinates are equivalent. On the contrary, the existence of a preferred chronology
means that (as far as causality is concerned) there exists some preferred class
of inertial coordinates, or equivalently some preferred rods and clocks.
This seems to be physically unacceptable.

As a consequence, one \textit{cannot} refer to causality in order
to assert that nothing can travel faster than one of these speeds.
Accordingly, causality does not require one of these metric to
define a wider cone than the others everywhere on spacetime.
Bimetric theories of gravity greatly enlightened this point, since
both the gravitational and the matter metric could be used to
define a ``natural'' chronology on spacetime. Depending on the
choice made, one then finds that nothing can travel faster than
gravity or light. These two opposite requirements can be found in
the literature.

On the contrary, our definition of causality in which no prior
chronology is assumed (by means of our ``mixed'' chronology, see
Sec.\ref{CausalityAndSuperluminalBehav}), enables a causal theory
to involve ``superluminal'' propagations. Actually the very notion
of superluminal behavior is no more meaningful in that framework,
and the only point is that some degrees of freedom can propagate
faster than other ones. Moreover the causal cones may even tip
over each other depending on the location on spacetime. As an
application, in k-essences theories, causality does \textit{not}
require the sign of $F''(X)$ to be fixed (and notably negative),
and, in bimetric theory causality does not require either the
gravitational light-cone to be wider than the matter one (nor
conversely).

What actually requires causality is first that the equations of
motion have a well-posed Cauchy problem. This strongly depends on
the precise form of the dynamics, and throughout this paper we
used classical results \cite{Wald} on that subject. We also
discussed in detail the fact that global properties of spacetimes
may break causality. We pointed out that this already occurs in
standard GR \textit{so that it cannot be related to some intrinsic
disease of superluminal propagation}. We discuss three very
similar cases: Gott's cosmic strings \cite{Gott}, the two bubbles
of non trivial vacua in k-essences theories \cite{Arkani-Hamed}
and the two Casimir experiments of \cite{LiberatiVisser}. What
thus became clear was that such non trivial global properties may
be actually suppressed by subtle boundaries effects. This would be
an illustration of the chronology protection conjecture.

\begin{acknowledgments}
The author is indebted to G. Esposito-Far\`ese for many useful
discussions and comments, and acknowledge discussions with C.
Deffayet, R. Durrer, E. Flanagan, J. Larena, J. Moffat and J.-P.
Uzan.
\end{acknowledgments}

\end{document}